# Giant frequency dependence of dynamic freezing in nanocrystalline ferromagnetic LaCo$_{0.5}$Mn$_{0.5}$O$_3$

R. Mahendiran[1,2], Y. Bréard[2], M. Hervieu[2], B. Raveau[2], and P. Schiffer[1]

[1]Department of Physics and Materials Research Institute, 104 Davey Laboratory, Pennsylvania State University, University Park, PA 16802 USA

[2]Laboratoire CRISMAT, ISMRA, Université de Caen, 6, Boulevard du Maréchal Juin, 14050 Caen Cedex, France

**Abstract**

We have investigated the magnetic properties of nanocrystalline LaCo$_{0.5}$Mn$_{0.5}$O$_3$. The temperature dependence of the imaginary part of the a.c. susceptibility shows a strongly frequency dependent maximum at a temperature, $T_f$, which is well below the ferromagnetic transition temperature ($T_C \sim 230$ K). The frequency dependence of $T_f$ obeys the Arrhenius relation, $f = f_o\exp(-E_a/k_BT)$, with physically reasonable values of $f_o = 10^9$ Hz and $E_a/k_B = 1518$ K. The frequency shift of $T_f$ per decade of frequency is one of the highest values observed in any magnetic system, and a similarly large value is also found in LaCo$_{0.4}$Mg$_{0.1}$Mn$_{0.5}$O$_3$, suggesting that such behavior is intrinsic despite the apparent presence of long range ferromagnetic order.

The perovskite manganites $La_{1-x}A_xMnO_3$ (A = Ca, Sr, Ba) have been a subject of intense research since the 1950s,[1,2,3] and there has been renewed recent interest due to the observation of "colossal magnetoresistance" near the transition to the double-exchange mediated ferromagnetic metallic state.[4] The Mn-site substituted materials of the form $LaMn_{1-x}M_xO_3$ (M = Co, Ni, Ga) are also ferromagnetic for x = 0.5.[5,6,7,8,9,10] Although the Mn, Co and Ni ions are in trivalent states in the ternary compounds $LaMnO_3$, $LaCoO_3$ and $LaNiO_3$, there is a tendency towards charge disproportionation in the quaternary materials $LaMn_{1-x}M_xO_3$ (i.e. a combination of $Mn^{4+}$ and $M^{2+}$),[9,9] and ferromagnetism in these insulating compounds is believed to be mediated by either vibronic superexchange interactions between $Mn^{3+}$ ions[5] or positive superexchange interactions between $Mn^{4+}$ and $M^{2+}$ ions (M = Co or Ni).[7] The Curie temperature in $LaMn_{1-x}M_xO_3$ series reaches a maximum for x = 0.5 ($T_C$ = 220-240 K for M = Co and $T_C$ = 280 K for M = Ni),[5-10] and there have also been suggestions of cationic ordering of $Mn^{4+}$ and $M^{2+}$ ions in these compositions.[9-10]

In this report, we investigate the a.c. and d.c. magnetic susceptibilities of ferromagnetic $LaCo_{0.5-x}Mg_xMn_{0.5}O_3$ ($x$ = 0, 0.1). We find a frequency dependent maximum in the imaginary part of the a.c. susceptibility at a temperature, $T_f$, well below the ferromagnetic ordering transition. This maximum indicates a dynamic spin freezing, which is surprising given the apparent presence of long range ferromagnetic order. Furthermore, the frequency dependence of $T_f$ is one of the strongest reported in any magnetic material, further indicating an unusual physical origin to this phenomenon.

Polycrystalline samples of LaMn$_{0.5-x}$Mg$_x$Co$_{0.5}$O$_3$ ($x$ = 0, 0.1) were prepared by the low temperature nitrate method as described by Nishimori *et al.*,[11] and the samples were characterized by x-ray diffraction, energy dispersive x-ray analysis, and electron microscopy (JEOL 2000). The a.c. susceptibility ($H_{ac}$ = 2 Oe rms) and the d.c. magnetization (*M*) were measured using a Quantum Design PPMS cryostat and a SQUID magnetometer, respectively.

In Fig. 1(a) we show a histogram of grain size distributions of LaCo$_{0.5}$Mn$_{0.5}$O$_3$ obtained with transmission electron microscopy. The low temperature synthesis resulted in small grains (40 - 160 nm) with an average grain size of about 70 nm in both materials. The electron microscopy for $x$ = 0 sample at 300 K revealed that the majority of the grains have orthorhombic structure with GdFeO$_3$ type distortions (space group *Pnma*) with [100] and [001] oriented domains due to twinning as commonly observed in many other La-site doped manganites. In these data, we also observe superlattice reflections along the [100]* direction at incommensurate positions (see two arrows in Fig. 1(b)) at room temperature in a small fraction of the grains (<5%). The amplitude of the modulation vector (*q*) which characterizes the superlattice reflections varies between $q$ = 0.42 and 0.44 in different grains, and no significant change in the value of *q* was found down to *T* = 92 K, the lowest temperature studied. This superstructure indicates some regions of short range ordering with dimension about 10 nm, which are also seen (with much weaker intensity) in the $x$ = 0.1 sample. A possible origin could be ionic ordering among the Co$^{2+}$ and Mn$^{4+}$ ions,[9-10] but a definitive identification of such ordering would require high resolution electron microscopy or neutron diffraction study in single crystal samples which are beyond the scope of the present work.

The main panel of Figure 2 shows the temperature dependence of the d.c. magnetic susceptibility ($M/H$) while warming from 5 K in a field of $H = 0.01$ T after zero field cooling (ZFC) and during cooling in the same field (FC) from 300 K. The sharp increase in the FC susceptibility around $T_C \approx 230$ K for $x = 0$ ($\approx 208$ K for $x = 0.1$) indicates the onset of ferromagnetic ordering. While the FC susceptibility continues to increase with decreasing temperature as expected for a conventional ferromagnet, the ZFC susceptibility deviates from the FC curve just below $T_C$, suggesting the importance of domain effects in this material. This is also indicated by the large coercive field ($H_C \approx 0.55$ T), which is demonstrated in figure 3 where we plot the field dependence of the magnetization, $M(H)$. Note also that the maximum value of $M = 2.7\mu_B$/f.u. for $x = 0$ ($M = 2.6\mu_B$/f.u. for $x = 0.1$) at $H = 7$ T. This saturation moment is close to the theoretical spin-only value of $3\mu_B$/f.u. for $x = 0$ ($2.85\mu_B$/f.u. for $x = 0.1$) expected for the ferromagnetic alignment of $Co^{2+}$ and $Mn^{4+}$ spins.[9] This nearly complete saturation of the moment is important in that it demonstrates the presence of long-range ferromagnetic order in this material for $T < T_C$.

The main panel of Fig. 4 shows the temperature dependence of the real part of the a.c. susceptibility ($\chi'$) at different frequencies for $x = 0$, and the inset shows corresponding data for $x = 0.1$. For both compounds, $\chi'$ increases rapidly at the onset of ferromagnetic order and decreases at lower temperatures. Although the change in the frequency affects the values of $\chi'$, there is no clear shift of the maximum at $T_C$ and only a small shift of $\chi'(T)$ for $T < T_C$. The imaginary part of the a.c. susceptibility, $\chi''(T)$, shows a rise at $T_C$ and a clear maximum well below $T_C$ at a temperature, $T_f$, as shown in figure 5. This lower temperature feature (which is seen for both samples) is strongly frequency dependent, shifting down in temperature, decreasing in magnitude, and broadening with decreasing frequency. There is also a small step like feature around $T$ = 175 K for $x$ = 0, seen clearly at low frequencies which is also reflected in the ZFC d.c. susceptibility (Fig. 2 inset). We speculate that this step is caused by the onset of magnetic ordering within the regions which show superstructure in the electron diffraction.

The appearance of the frequency dependent maximum in $\chi''(T)$ below $T_C$ suggests dynamic spin freezing at a temperature $T_f < T_C$ within the ferromagnetic state. The absence of a corresponding feature in $\chi'(T)$, can be understood as a consequence of the large real part of the susceptibility associated with ferromagnetism.[14] To characterize the frequency dependence of this feature, we plot $1/T_f$ versus $\ln(f)$ in Fig. 6. The observed linear behavior in this plot implies that the dynamic spin freezing follows the Arrhenius law, $f = f_0\exp(-E_a/k_BT_f)$, with $\tau_0 = 1/f_0 = 4.14\times10^{-9}$ s and $E_a/k_B$ = 1518 K for $x$ = 0 and $\tau_0$ = $1.31\times10^{-9}$ s and $E_a/k_B$ = 1923 K for $x$ = 0.1. The observed values of $\tau_0$ and $E_a/k_B$ are physically reasonable and the observed $\tau_0$ is in the range expected for superparamagnetic particles ($\tau_0 = 10^{-8}$-$10^{-13}$ sec).[15] The frequency dependence of $T_f$ can be quantified by $g = \Delta T_f/[T_f\Delta(\log_{10}[f])]$ = 0.23 for $x$ = 0 ($g$ = 0.19 for $x$ = 0.1), which is one of the largest values reported in any magnetic material. The extremely large $g$ found in both samples suggests that this is an intrinsic behavior. By contrast, typical values for spin glasses are $g$ ~ 0.005-0.01,[12] the rare-earth-site doped manganites exhibit $g$ < 0.05,[13] and $g$ ~ 0.03 – 0.06 was reported in the 'cluster glass' compound $La_{0.5}Sr_{0.5}CoO_3$.[14] Higher values of $g$, of order 0.06-0.09, are found in Fe nanograins embedded in amorphous $Al_2O_3$ and $Fe_2O_3$ particles dispersed in polymer.[15] Freezing with $g \geq 0.1$ occurs in a few systems with superparamagnetic-blocking-like transitions in materials such as $Ca_3CoRhO_6$ ($g$ = 0.10)[16],

La$_{0.994}$Gd$_{0.06}$Al$_2$ ($g$ = 0.13)[17], Ni vermiculite intercalation compound ($g$ = 0.24) [18], or Ho$_2$O$_3$. B$_2$O$_3$ ($g$ = 0.28)[19], molecular clusters of Mn-12 ($g$ = 0.24)[20] or in the exotic glassy freezing of the spin ice material Dy$_2$Ti$_2$O$_7$ ($g$ = 0.18)[21]. In all of these compounds with $g$ > 0.1, $T_f \lesssim$ 10 K, which contrasts sharply with the maximum $T_f \gtrsim$ 100 K in the present study. Furthermore, none of the above compounds with $g$ > 0.1 also exhibit long range magnetic order.

The observed large values of $g$ and physical reasonable value of $\tau_o$ ( ~ 10$^{-9}$ sec) suggest superparamagnetic like relaxation could be responsible for the dynamic spin freezing in our compounds. As mentioned above, such a giant frequency dependence is seen in superparamagnetic materials in which magnetic particles are well-separated and can respond independently to the applied a.c. magnetic field. Such behavior is not a good model for the present case, however, since this material is a long range ordered ferromagnet. We believe that there are two possible explanations for the observed dynamic spin freezing at $T_f$. First, the superstructured, ionic ordered domains which are non magnetic at room temperature could become magnetic at low temperatures (the anomaly observed around $T$ = 175 K in d.c. and a.c. susceptibilities for $x$ = 0 is possibly caused by such ordering). Since the local symmetry of these domains is different from the rest of the matrix, the relaxation of magnetization within these domains could be independent of the rest of the matrix, and may be analogous to dynamical freezing of nanoscale size superparaelectric domains found in relaxor ferroelectrics.[22] Another possibility is that the peak in $\chi''(T)$ results from oscillations of pinned domain walls.[23] The nanometer sized grains in this material will certainly affect the domain wall structure and pinning and thus could lead to the observed anomalously large values of $g$ (which are not observed in other signatures of domain wall effects in $\chi''(T)$[23]).

Regardless of the origin of the large $g$, the observed dynamic magnetic freezing phenomenon is qualitatively different from previously observed behavior in magnetic oxides. The existence of glassy behavior in a long-range-ordered magnetic material, provides a further indication of the rich physics accessible in nanometer-scale magnetic materials. Further study with local probes such as Mössbauer spectroscopy, muon spin relaxation or small angle neutron scattering will be important in further investigating the microscopic origins of the behavior.

Acknowledgment: The work at Pennsylvania State University was supported by NSF grant DMR-0101318. R. M. also acknowledges financial support form MENRT (France). We are also grateful for helpful discussions with J. Blasco.

**Figure Captions**

1. (top) Histogram of grain size of $LaCo_{0.5}Mn_{0.5}O_3$ obtained from the transmission electron microscopy. The average grain size is 70 nm. (bottom) Electron diffraction image of the superlattice reflections (shown by two arrows) along [100]* direction in $LaCo_{0.5}Mn_{0.5}O_3$ at room temperature.

2. Temperature dependence of zero field cooled (open symbols) and field cooled (closed symbols) d.c. magnetic susceptibilities (M/H) of $LaCo_{0.5-x}Mg_xMn_{0.5}O_3$ for $x = 0$ and $x = 0.1$ at $H = 0.01$ T. The inset shows a presence of a weak anomaly around $T = 175$ K in the zero field cooled magnetization of $x = 0$.

3. The field dependence of the magnetization of $LaCo_{0.5-x}Mg_xMn_{0.5}O_3$ for $x = 0$ and $0.1$ at $T = 5$ K. The observed magnetization at $H = 7$ T is close to the saturation moment (see text for details).

4. Temperature dependence of the real part ($\chi'$) of the a.c. susceptibility at different frequencies for $LaCo_{0.5}Mn_{0.5}O_3$. The arrow indicates increasing frequency ($f = 10, 100, 1,000, 2,000,$ and $10,000$ Hz). Inset shows the data for $x = 0.1$.

**5.** Temperature dependence of the imaginary ($\chi''$) part of the a.c. susceptibility at different frequencies for (a) $x = 0$ and (b) $x = 0.1$. The appearance of a frequency dependent maximum at $T_f$ is indicated by the arrows.

**6.** The frequency dependence of the maximum in $\chi''(T)$ data for $x = 0$ and $x = 0.1$. The solid lines are fit to Arrhenius law $f = f_0 \exp(-E_a/k_B T)$ as described in the text.

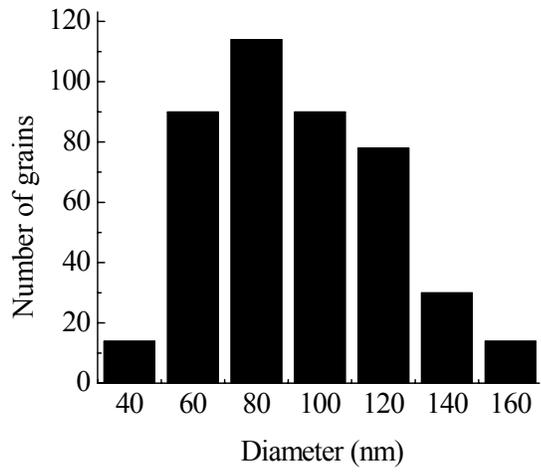

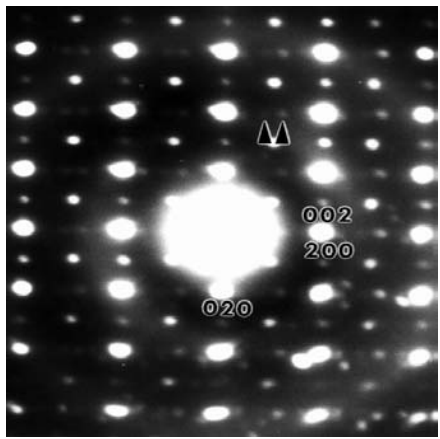

**FIG. 1**

R. Mahendiran *et al.*

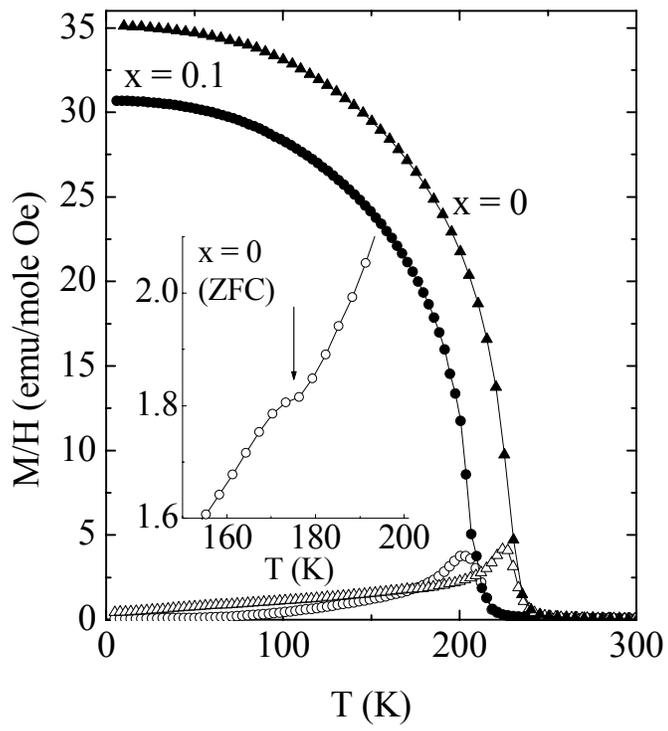

Fig. 2
R. Mahendiran et al.

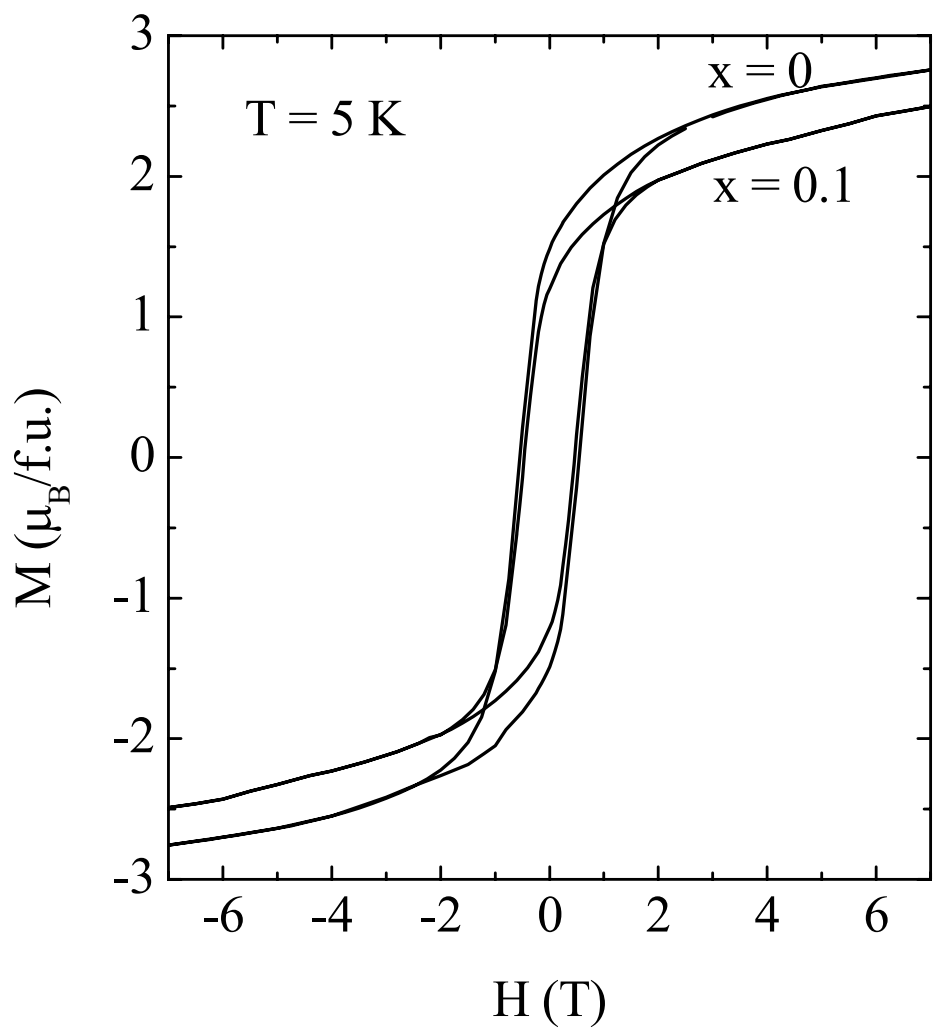

Fig. 3
R. Mahendiran et al.

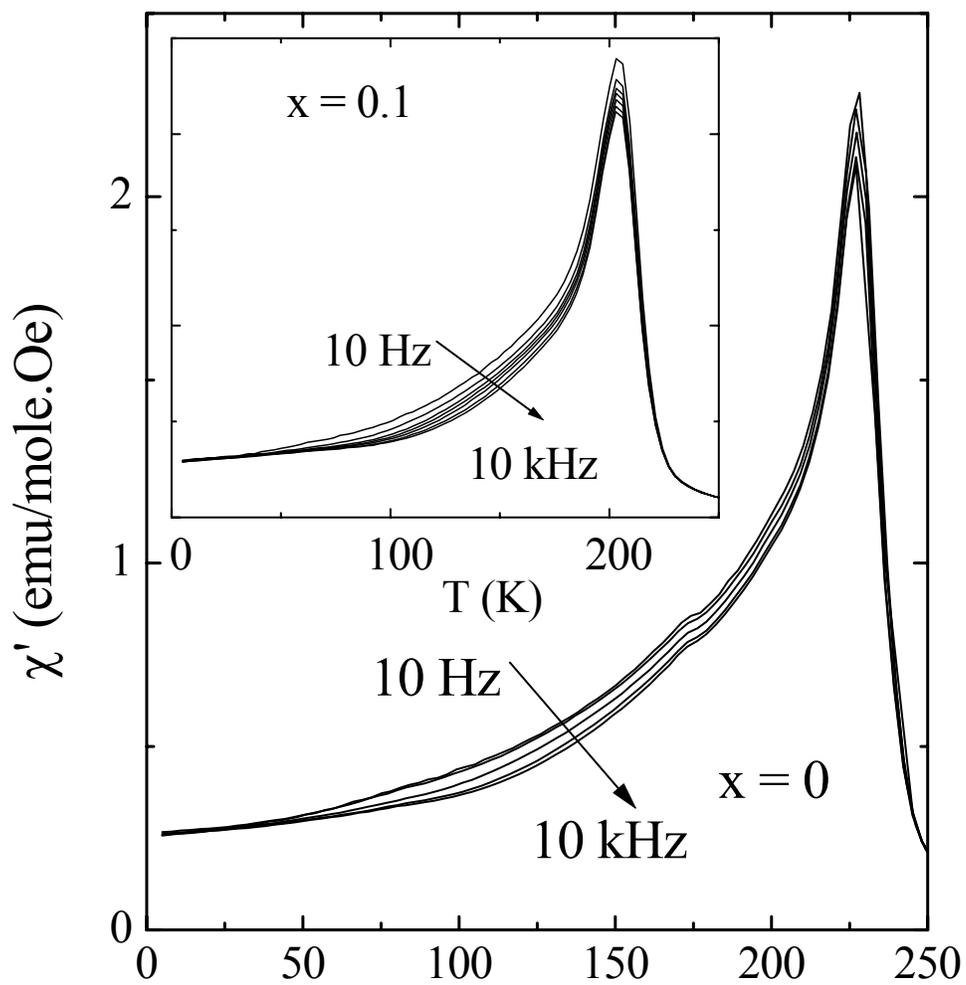

FIG. 4
R. Mahendiran et al.

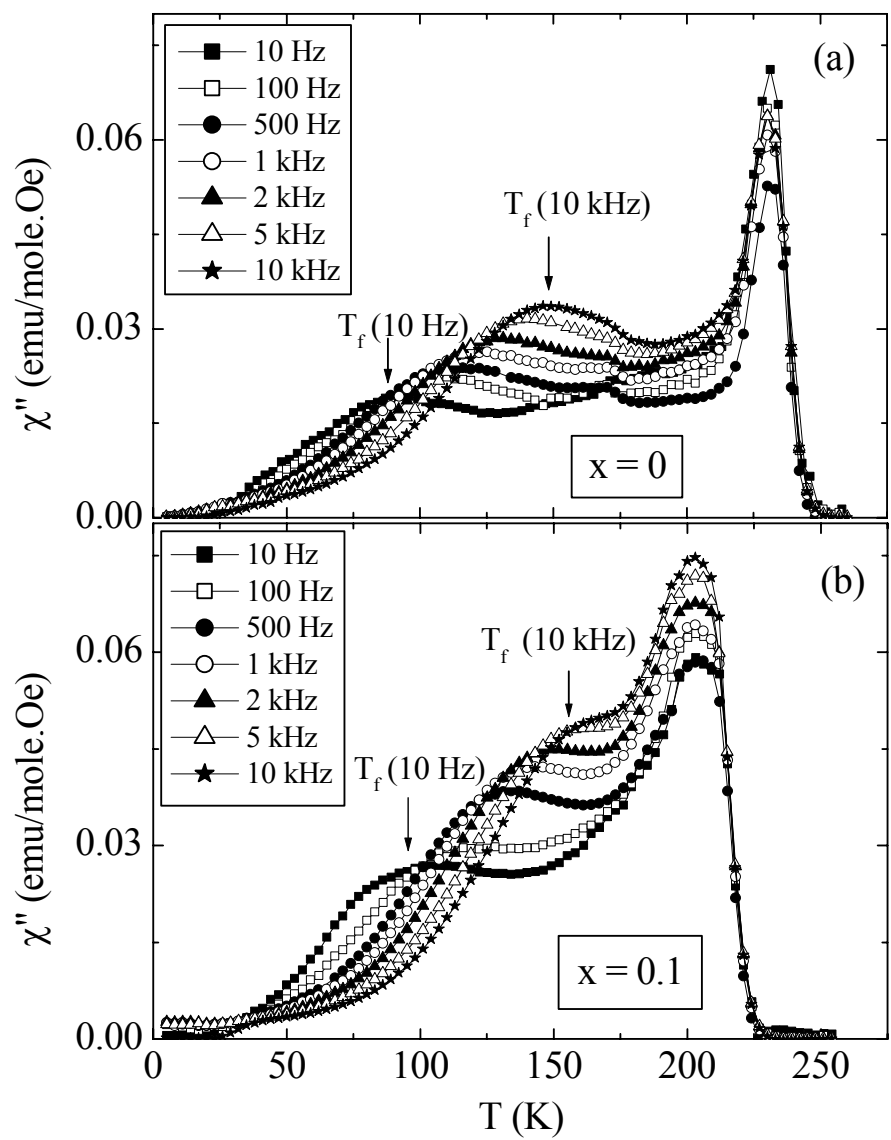

FIG. 5
R. Mahendiran et al.

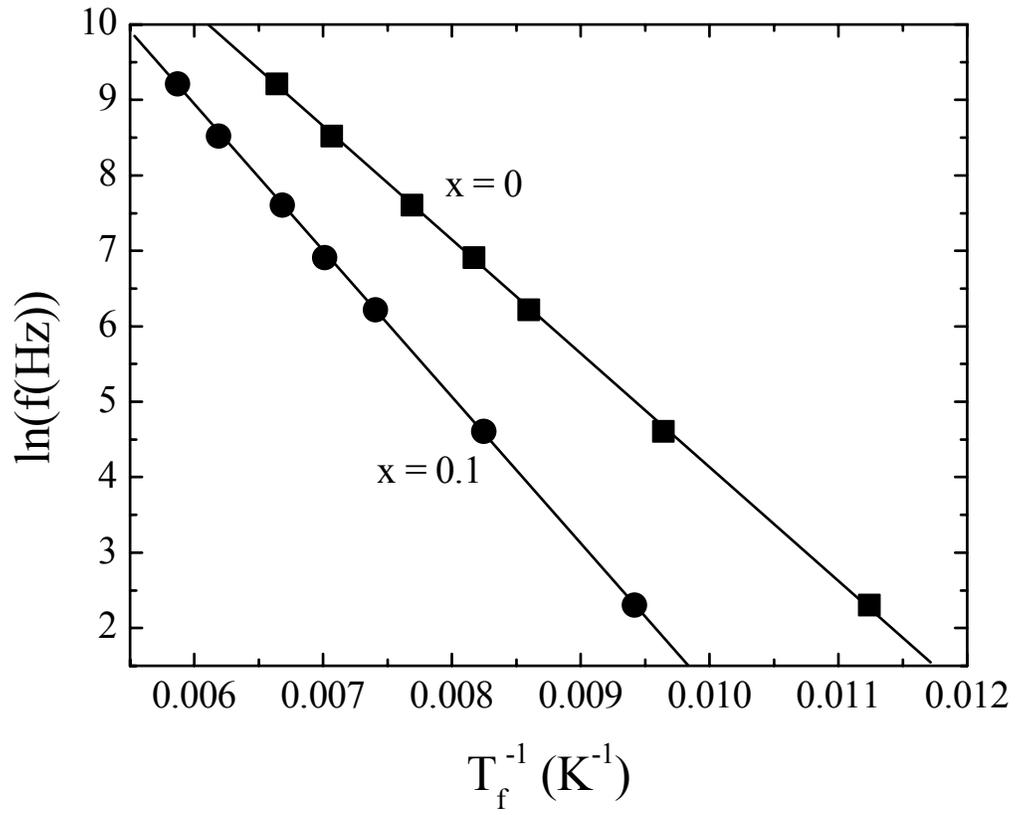

FIG. 6
R. Mahendiran et al.